# Nonrelativistic and nonmagnetic control of terahertz charge currents via electrical anisotropy in RuO$_2$ and IrO$_2$

Sheng Zhang[1†], Yongwei Cui[1,2†], Shunjia Wang[1†], Haoran Chen[1,2], Yaxin Liu[1], Wentao Qin[1,2], Tongyang Guan[1], Chuanshan Tian[1], Zhe Yuan[3], Lei Zhou[1], Yizheng Wu[1,2*], Zhensheng Tao[1*]

[1] State Key Laboratory of Surface Physics and Department of Physics and Key Laboratory of Micro and Nano Photonic Structures (MOE), Fudan University, Shanghai 200433, China

[2] Shanghai Research Center for Quantum Sciences, Shanghai, China

[3] Center for Advanced Quantum Studies and Department of Physics, Beijing Normal University, Beijing 100875, China

†These authors contributed equally to this work.

*Yizheng Wu, wuyizheng@fudan.edu.cn;

*Zhensheng Tao, ZhenshengTao@fudan.edu.cn

**Abstract**

Precise and ultrafast control over photo-induced charge currents across nanoscale interfaces could lead to important applications in energy harvesting, ultrafast electronics, and coherent terahertz sources. Recent studies have shown that several relativistic mechanisms, including inverse spin-Hall effect[1,2], inverse Rashba-Edelstein effect[3,4] and inverse spin-orbit-torque effect[5], can convert longitudinally injected spin-polarized currents from magnetic materials to transverse charge currents, thereby harnessing these currents for terahertz generation. However, these mechanisms typically require external magnetic fields and suffer from low spin-polarization rates and low efficiencies of relativistic spin-to-charge conversion. In this work, we present a novel nonrelativistic and nonmagnetic mechanism that directly utilizes the photo-excited high-density charge currents across the interface. We demonstrate that the electrical anisotropy of conductive oxides RuO$_2$ and IrO$_2$ can effectively deflect injected charge currents to the transverse direction, resulting in efficient and broadband terahertz radiation. Importantly, this new mechanism has the potential to offer much higher conversion efficiency compared to previous methods, as conductive materials with large electrical anisotropy are readily available, whereas further increasing the spin-Hall angle of heavy-metal materials would be challenging. Our new findings offer exciting possibilities for directly utilizing these photo-excited high-density currents across metallic interfaces for ultrafast electronics and terahertz spectroscopy.

Precise control of charge-carrier transport across nanoscale interfaces at ultrafast speeds is essential for the advancement of various modern technologies, including solar cells[6], photosysthesis[7], and high-efficiency optoelectronic devices[8]. It has been demonstrated that when metallic interfaces are excited by strong femtosecond laser pulses, enormous current density exceeding $10^{10}$ A cm$^{-2}$ can be produced[1,9]. This current density is several orders of magnitude higher than those typically used in electronic devices. If harnessed, these high-frequency and high-density charge currents could revolutionize the field of ultrafast electronics[10] and lead to the development of high-speed storage devices[11,12] and coherent terahertz sources[2]. However, due to the buried nature and nanometer scale of these interfaces, it remains challenging to utilize the laser-induced high-density charge currents in practical applications.

Recently, Kampfrath *et al.* demonstrated a promising approach for utilizing these currents by combining a heavy-metal (HM) thin film, such as Pt, with a ferromagnetic (FM) thin film[1]. By exploiting the relativistic inverse spin-Hall effect (ISHE), longitudinal spin-polarized currents injected from the FM layer can be deflected to transverse charge currents in the HM layer, generating strong and broadband terahertz radiation[2,13,14]. In this context, longitudinal currents are defined as those flowing perpendicular to the interface. Metasurface-structured devices have also been demonstrated recently for simultaneous generation and manipulation of terahertz waveforms[15,16]. Moreover, this concept has attracted great interest because it provides an all-optical, contact-free method for probing the transient state of magnetism with sub-picosecond time resolution in FM materials[17–19], antiferromagnetic materials[20], and for reliably measuring the spin-Hall angle of HM materials[21]. Later, other relativistic mechanisms, including inverse Rashba-Edelstein effect[3,4] and inverse spin-orbit-torque effect[5], are also found capable of spin-to-charge conversion. The former has comparable efficiency with the ISHE, while the latter effect is much weaker.

All the current-deflection mechanisms described above rely on a two-step process involving generation of spin-polarized currents and relativistic spin-to-charge conversion. The spin-polarized currents are typically extracted from the laser-induced charge currents through super-diffusive spin scattering[22,23], resulting in a spin polarization rate of 0.2 to 0.4 within the spin diffusion length[24,25]. As a result, an external magnetic field is usually required to saturate the magnetization of the FM materials, although field-free emitters have recently been realized by utilizing exchange bias between antiferromagnetic and FM nanofilms[26]. In the second step, the efficiency of the relativistic spin-to-charge conversion is characterized by the spin-Hall angle $\gamma$. For Pt, which is known for its strong spin-orbit coupling, $\gamma$ is typically around $0.1^5$, while the conversion efficiency of an AgBi interface is estimated to be between 0.064 and $0.16^3$. Consequently, the ability to fully utilize the interface transient currents is hindered by the low conversion efficiencies in these two steps. The conversion efficiency could be significantly improved if one could directly and efficiently control the laser-induced charge currents across the interface, rather than relying on spin-polarized currents.

In this work, we report a novel nonrelativistic and nonmagnetic mechanism for direct conversion of laser-excited high-density longitudinal charge currents to transverse ones, leading to efficient terahertz-wave generation without the need for external fields. The generation process is initiated by the super-diffusive charge current injected from the adjacency of an optically excited metal thin film, which is then deflected from the longitudinally injected direction to the transverse direction by the anisotropic electrical conductivity of the conductive rutile oxides $RuO_2$ and $IrO_2$. Notably, $RuO_2$ is recently found to be an itinerant antiferromagnetic material[27,28] that has attracted enormous interest in magneto-electronic research[29–35], while $IrO_2$ is nonmagnetic[36]. Our results show that the terahertz emission is highly sensitive to the crystal orientation but not influenced by the polarization of the excitation laser. This distinguishes our mechanism from the aforementioned magnetic near-infrared (NIR)-to-terahertz conversion mechanisms[1,3–5] that rely on relativistic spin-orbit coupling, as well as from other nonmagnetic mechanisms such as optical rectification[37,38] or difference-frequency generation[39,40] that require coherent wave mixing of the excitation laser. The conversion efficiency of the $IrO_2$ sample matches that of the ISHE, and this new mechanism can potentially further improve the efficiency by implementing conductive materials with stronger electrical anisotropy. These new findings open up new possibilities to directly harness interface high-density charge currents for ultrafast electronics and terahertz spectroscopy.

Figure 1a illustrates the schematic of the experimental setup. The device is based on a heterostructure composed of a single-crystal film of either $RuO_2$ or $IrO_2$ and a nonmagnetic metal (NM) thin film, both of which are nanometers thick. Several different metals (Cu, Pt, W, Ir) were used for the NM layer. The $RuO_2$ and $IrO_2$ films are both conductive rutile oxide belonging to the space group P42/mnm with unequal lattice parameters ($a=b>c$, see Fig. 1b). As a result, they are both electrically anisotropic conductors (EAC) with $\sigma_\parallel < \sigma_\perp$, where $\sigma_\parallel$ and $\sigma_\perp$ are the conductivity along the *c*-axis and that in the *a-b* plane, respectively. The single-crystal $RuO_2$ or $IrO_2$ film was deposited on the $TiO_2$ or $Al_2O_3$ substrates and then capped by a NM film. See Methods and Supplementary Materials (SM) Section S1 for the details of sample preparation and basic characterizations.

In our experiment, the NM/EAC heterostructure is excited by femtosecond pulses (duration of ~25 fs, center wavelength 1030 nm, repetition rate 100 kHz). The excitation laser pulses are generated through high-quality pulse compression enabled by solitary beam propagation in periodic layered Kerr media[41,42]. The laser beam is incident normally from the substrate side onto the heterostructure along the -z direction, and the beam radius is focused to around 0.7 mm on the sample. The x-y-z coordinates refer to the laboratory frame, while $a$, $b$ and $c$ are the crystallographic axes. The polarization of the excitation laser can be adjusted with a combination of waveplates to be either linearly polarized with a polarization angle $\alpha$ in the x-y plane, or to be circularly polarized. The sample temperature can be changed between 77K and 500 K. In the experiment, the orientation of the $RuO_2$ or $IrO_2$ crystals is varied by growing the sample on different substrates or by rotating the sample in plane around $z$. Hence, we define the polar angle $\theta$ between the crystal $c$-axis and the $z$-axis, and the azimuthal angle $\varphi$ between the projection of the $c$-axis in the x-y plane and the x-axis (see Fig.1a and b). Finally, the emitted terahertz waveforms are detected using a polarization- and time-resolved terahertz spectroscopy setup based on electro-optic sampling (EOS)[43–45]. The waveforms of the two orthogonal terahertz polarizations ($E_x$ and $E_y$) can be resolved (see Methods).

Figure 1b presents the terahertz waveforms generated at room temperature by a $RuO_2$(10 nm) thin film, Pt(2 nm)/$RuO_2$(10 nm) and Pt(2 nm)/$IrO_2$(10 nm) heterostructures, in comparison with a spintronic terahertz emitter composed of a Pt(2 nm)/Fe(2 nm) heterostructure whose conversion efficiency has been optimized[14]. The signal from the spintronic emitter was measured under an external magnetic field that magnetizes the FM layer, while those from the RuO$_2$ thin film and from the NM/EAC heterostructures were measured without any external fields. In this measurement, the RuO$_2$ and IrO$_2$ films are both (101)-oriented, which are grown on the TiO$_2$(101) substrates.

First of all, we find that capping RuO$_2$ with a 2 nm Pt layer enhances the emitted terahertz amplitude by a factor of 10, indicating the strong influence of the heterostructure on the NIR-to-terahertz conversion. The Pt/IrO$_2$ heterostructure can deliver terahertz amplitude about 3 times as strong as Pt/RuO$_2$. Remarkably, the terahertz emission from Pt/IrO$_2$ is almost as strong as that from Pt/Fe, indicating high conversion efficiency. The strength of this terahertz signal is comparable to those generated by several commercial terahertz sources based on nonlinear optical crystals and photoconductive switches[2,14,16]. We also find that the terahertz spectra of different samples are almost identical, as shown in Fig. 1c.

Previous studies have shown that the ISHE in the NM layer can cause deflection of spin-polarized currents, leading to significant enhancement of terahertz generation in heterostructures, such as those involving a laser-excited ferromagnetic layer[1,2,13,14] and a ferrimagnetic yttrium iron garnet (YIG) layer driven by spin-Seebeck effect[17], or an antiferromagnetic NiO layer with coherently excited spin currents[20]. The polarity and amplitude of the emitted terahertz field are dependent on the spin-Hall angle ($\gamma$) of the NM material. In Fig. 2a, we investigate the influence of different NM materials on terahertz emission from NM/EAC heterostructures. The terahertz-wave amplitudes, the spin-Hall angles ($\gamma$), and the optical absorption coefficients (OAC) of different NM materials are summarized in Fig. 2b. Here, the most important observation is that the terahertz-wave polarity from the W/RuO$_2$ heterostructure is not reversed, despite $\gamma$ of W is of opposite sign compared to that of Pt[2]. This behavior contrasts with the spintronic emitter (see SM Section S2). Further, we find that even though Cu has a small spin-Hall angle, the terahertz amplitude from Cu/RuO$_2$ is comparable to that from Pt/RuO$_2$ (Fig. 2b). These results therefore clearly distinguish the conversion mechanism of the NM/EAC heterostructures from ISHE.

In Fig. 2c and d, the dependence of terahertz emission on the polarization states of the excitation laser is further investigated. The RuO$_2$ crystal used in this measurement is (101) oriented with its $c$-axis fixed in the $x$-$z$ plane at $\varphi=0°$ (Fig. 1a). The results show that the emitted terahertz wave maintains a constant field amplitude and linear polarization along $x$, when the polarization angle ($\alpha$) of the linearly polarized excitation laser is varied (Fig. 2c). Note that this result is obtained after correcting for the birefringent effect of the TiO$_2$ substrate. The polarization-independent result is further confirmed by samples grown on Al$_2$O$_3$($1\bar{1}02$) substrate where the optical birefringent effect is not present (see SM Section S4). The $E_x$-$E_y$ projections of the terahertz waves under different $\alpha$ are shown in the inset of Fig. 2c. Furthermore, there is almost no difference in the amplitudes or waveforms of the terahertz signals excited by linearly and circularly polarized laser pulses (Fig. 2d). Similar results can be obtained from the Pt/IrO$_2$(101) heterostructures (see SM Section S8), leading to the conclusion that the terahertz generation from the NM/EAC emitters are not affected by the polarization of the excitation laser.

It should be noted that the independence of terahertz emission on the excitation-laser polarization rules out optical rectification[37,38] in the RuO$_2$ or IrO$_2$ crystals as the mechanism for the NIR-to-terahertz conversion. This result is also in distinction from the recent Pt/NiO emitter[20], where the coherent spin-current generation in NiO depends strongly on the laser polarization. Instead, the polarization independence here is in line with the spintronic emitters, where the terahertz emission is initiated by the incoherent conversion of optical energy to charge/spin currents[1,2]. This is further supported by the almost identical terahertz spectra from the two different types of emitters (see Fig. 1c), indicating similar carrier dynamics.

Nonetheless, our results strongly indicate that the efficient NIR-to-terahertz conversion in the NM/EAC heterostructures is nonmagnetic in origin. Firstly, ISHE has been excluded. We also find that the emitted terahertz waves are unaffected by external magnetic fields (see SM Section S5). Secondly, while $RuO_2$ is an itinerant antiferromagnetic material[27,28], $IrO_2$ is known to be nonmagnetic[36]. Nonetheless, the emission properties from the two heterostructures are very similar (see SM Section S8). Thirdly, the temperature-dependent results show that the terahertz-wave amplitude from Pt/$RuO_2$(101) increases monotonically up to 500 K (see SM Section S6), which is higher than the reported Néel temperature of $RuO_2$ thin films[28].

Since the optical absorption coefficients of $RuO_2$ and $IrO_2$ are more than one order of magnitude smaller than that of the NM materials at the wavelength of ~1 μm[46], we believe that the terahertz emission originates from the injection of super-diffusive charge currents[22,23] from the optically excited NM layer into the $RuO_2$ or $IrO_2$ crystals (see inset of Fig. 1a). This is supported by the fact that the general trend of the terahertz amplitude vs. NM materials is in semiquantitative agreement with the optical absorption coefficients of the NM materials (Fig. 2b). Here, our results also suggest that the thermally driven Seebeck effect is unlikely to be responsible for the charge-current injection. This is because the terahertz waveforms emitted from devices with different NM materials are almost identical, whereas distinctive temperature dynamics would be expected in these metals after laser-pulse excitation (see SM Section S3).

The observation of the *x*-polarized terahertz field indicates the existence of a transverse charge current flowing along the *x* direction in our experimental setup (Fig. 1a). We find that this can be attributed to the anisotropic electrical conductivity of single-crystal RuO$_2$ and IrO$_2$. Due to the unequal lattice parameters, the second-order conductivity tensor is given by $\begin{pmatrix} \sigma_\perp & 0 & 0 \\ 0 & \sigma_\perp & 0 \\ 0 & 0 & \sigma_\parallel \end{pmatrix}$ in the crystal coordinate (*a-b-c*), where $\sigma_\parallel < \sigma_\perp$. By rotating the crystal under the azimuth angle $\varphi$ and the polar angle $\theta$, off-diagonal tensor components appear in the laboratory coordinate: $\sigma_{xz} = (\sigma_\perp - \sigma_\parallel) \cos\theta \sin\theta \cos\varphi$ and $\sigma_{yz} = (\sigma_\perp - \sigma_\parallel) \cos\theta \sin\theta \sin\varphi$, and the diagonal component $\sigma_{zz}$ becomes $\sigma_{zz} = \sigma_\perp \sin^2\theta + \sigma_\parallel \cos^2\theta$ (see Methods). As a result, when the charge current ($j_z$) is injected along -*z* (electron current along *z*), the conductivity anisotropy leads to the transverse charge current density of $j_x = j_z \beta_0 \cos\varphi$ and $j_y = j_z \beta_0 \sin\varphi$, where the coefficient of conductivity anisotropy $\beta_0$ is given by $\beta_0 \approx \frac{(\sigma_\perp - \sigma_\parallel) \cos\theta \sin\theta}{\sigma_\perp \sin^2\theta + \sigma_\parallel \cos^2\theta}$ when $(\sigma_\perp - \sigma_\parallel) \cos\theta \sin\theta \ll \sigma_\perp \sin^2\theta + \sigma_\parallel \cos^2\theta$. The amplitude of the *x*(*y*)-polarized terahertz field, $E_{x(y)}$, is proportional to the transverse currents, $j_{x(y)}$.

The above theory is confirmed by the experimental measurements under different crystal orientations [(101), (110), (100) and (001)], as shown in Fig. 3a. We find that only when the crystal orientation is (101) with $\theta=34.7°$ can strong terahertz emission be observed, and for $\varphi=0°$, only the *x*-polarized terahertz field is observed, because $\beta_0 \cos\varphi \neq 0$ and $\beta_0 \sin\varphi = 0$. On the other hand, when the *c*-axis is either aligned with *z* [(001) with $\theta=0°$] or in the *x-y* plane [(100) or (110) with $\theta=90°$], the terahertz emission in both polarizations is strongly suppressed, because $\beta_0 = 0$ under these conditions. Our results also show that, when the crystal orientation is (101), the polarization of the emitted terahertz field rotates following the azimuthal angle $\varphi$ (Fig. 3b). The peak amplitudes are summarized in Fig. 3c, and the sinusoidal behaviors of the $E_x$ and $E_y$ amplitudes are in excellent agreement with $\beta_0 \cos\varphi$ and $\beta_0 \sin\varphi$, respectively.

The ability to convert $j_z$ to $j_{x,y}$ is characterized by $\beta_0$ of different materials, which is analogous in position to the spin-Hall angle $\gamma$ within the ISHE formalism[1,47,48]. In Table 1, we list the experimentally measured $\sigma_\parallel$ and $\sigma_\perp$ of $RuO_2$ and $IrO_2$, respectively (see Methods). When the crystal orientation is (101), we find that $\beta_0$ of $IrO_2$ is ~5 times that of $RuO_2$. In Fig. 4a, we show that terahertz amplitudes from the Pt/$IrO_2$ and Pt/$RuO_2$ structures both grow linearly as a function of the incident pump fluence ($F$) when $F$<0.4 mJ cm$^{-2}$. Remarkably, the slope of the linear increase of Pt/$IrO_2$ is ~4.8 times that of Pt/$RuO_2$, in excellent agreement with the ratio of $\beta_0$ between the two materials. We note that, due to the low conductivity of $RuO_2$ and $IrO_2$, the impedance shunt effect only contributes approximately 15% of the difference in terahertz amplitudes (see Methods). The NIR-to-terahertz conversion efficiency of the Pt/$IrO_2$ heterostructure almost reaches that of Pt/Fe heterostructure. More interestingly, the signal increases of these two structures both deviate from a linear increase at $F$≈0.4 mJ cm$^{-2}$, while that from the Pt/$RuO_2$ structure continues to increase linearly at high pump fluence. This behavior may be attributed to the contrasting temperature-dependent behaviors of the three structures: When the laser excitation increases the sample temperature, the terahertz signal from Pt/$RuO_2$ increases monotonically with the rising sample temperature up to 500K, whereas those from Pt/$IrO_2$ and Pt/Fe structures both decrease (see SM Section S6).

In Fig. 4b and c, we plot the dependence of the terahertz amplitudes as a function of the thickness of the EAC layer ($d_{EAC}$) and the NM layer ($d_{NM}$), respectively. Here, we take the Pt/RuO$_2$ device as an example. Importantly, we find that the terahertz amplitude gradually increases as $d_{NM}$ increases from 0, and peaks at approximately 2 nm. This indicates that the deflection of the charge currents is not caused by the interface effect. In contrast, the terahertz amplitude as a function of $d_{EAC}$ exhibits much slower variation, and the maximum terahertz signal is generated when $d_{EAC} \approx 7.5$ nm.

Quantitatively, when we fix the crystal orientation with $\theta=34.7°$ and $\varphi=0°$, the amplitude of the x-polarized terahertz field ($E_x$) is directly related to the z-integration of the transverse charge current density ($\beta_0 j_z$) by[2]

$$E_x(\omega) = Z(\omega)e \int_0^{d_{EAC}} dz\, \beta_0 j_z(z, \omega), \qquad (1)$$

where $e$ is the elementary charge. Here, $Z(\omega)$ is the effective impedance of the heterostructure in the transverse direction shunted by the adjacent substrate and air spaces, which is related to the thickness of the EAC layer ($d_{EAC}$) and the NM layer ($d_{NM}$) (see Methods). The longitudinal current density $j_z$ is proportional to the density of the absorbed photons: $j_z \propto \frac{F_{abs}}{d_{NM}\, \hbar\omega_0}$, where $F_{abs}$ is the absorbed laser fluence of the NM layer and $\hbar\omega_0$ is the excitation photon energy. In addition, the spatial distribution of $j_z$ that contributes to the terahertz radiation is localized near the heterostructure interface, due to the finite hot-electron velocity-relaxation lengths in the EAC layer ($\lambda_{EAC}$) and the NM layer ($\lambda_{NM}$) (see Methods). The best fit to the experimental results are shown in Fig. 4b and c (solid lines), which yields $\lambda_{RuO_2} \approx 3.2$ nm and $\lambda_{Pt} \approx 1$ nm for RuO$_2$ and Pt, respectively. The latter is in consistent with previous work[2].

**Discussion and conclusion**

Our study has demonstrated a novel nonmagnetic and nonrelativistic mechanism for generating strong terahertz-wave emission by directly harnessing laser-excited charge currents across nanoscale interfaces. This approach utilizes the anisotropic electrical conductivity of materials and eliminates the need for conversion of charge currents to spin-polarized currents. Our results also highlight the importance of using conductive materials to enable efficient injection of laser-induced currents into the EAC layer. This is supported by the fact that the Pt/TiO$_2$(101) heterostructure does not generate terahertz radiation (Fig. 4c), although TiO$_2$(101) is an insulator exhibiting similar crystal anisotropy.

Compared to the ISHE mechanism, this novel mechanism could offer much higher conversion efficiency by selecting conductive materials with large electrical anisotropy, whereas further increasing the spin-Hall angle $\gamma$ of HM materials would be difficult. For example, the conductivity in the basal plane of a graphite thin layer is $\sigma_\perp \approx 10^6$ $\Omega^{-1}$ m$^{-1}$, while that normal to the plane is $\sigma_\parallel \approx 50$ $\Omega^{-1}$ m$^{-1}$ [49]. When charge currents are injected with an angle of $\theta$=0.5°-2° relative to the material normal axis, the significant difference in conductivity could possibly lead to terahertz emission with an order of magnitude higher intensity compared to that from the Pt/IrO$_2$(101) device in this study (see SM Section S7).

**Methods**

**1. Experimental details.**

Single-crystal $RuO_2$ and $IrO_2$ films were epitaxially grown on the double-polished $TiO_2$ or $Al_2O_3$ substrates by dc magnetron sputtering at 500 °C in a chamber with the base pressure better than than $2\times10^{-8}$ Torr. Both $TiO_2$ and $Al_2O_3$ substrates were pre-annealed at 500 °C for one hour before sample growth. Both $RuO_2$ and $IrO_2$ films were grown by reaction sputtering in the mixed atmosphere of Ar and $O_2$ with the ratio of 4:1. The normal metals Pt, W and Cu were deposited by dc magnetron sputtering at room temperature.

In the terahertz experiment, we excited the sample with femtosecond pulses (duration, 25 fs; center wavelength, 1030 nm; pulse energy, 15 µJ; repetition rate, 100 kHz; beam radius at the sample, 0.7 mm) under normal incidence from the substrate side. The terahertz electric field is subsequently detected by EOS using a 300 µm-thick (110)-oriented GaP crystal, with the two orthogonal components ($E_x$ and $E_y$) resolved using a broadband wire-grid polarizer. All the measurements were performed in a dry air atmosphere. The details of the polarization-resolved EOS setup can be found elsewhere[16].

For the electrical measurements, the single-crystal $RuO_2(100)$ and $IrO_2(100)$ films were patterned into devices with two orthogonal Hall bars through standard photolithography and Ar-ion etching. The current can flow through either the crystal *a* axis or *b* axis. The width and the distance between the two electrodes of the Hall bars are 150 μm and 600 μm, respectively. The electrical measurements were carried on a cryogenic probe station (LakeShore EMPX-HF) at room temperature. A dc current of 1 mA was injected into the longitudinal bar, and the voltage were detected by Keithley's 2182A nanovoltmeter.

## 2. Electrical anisotropic conductivity tensor.

Both RuO$_2$ and IrO$_2$ are rutile oxides with the P42/mnm space group, where Ru/Ir atoms occupy the center of stretched oxygen octahedrons. The conductivity tensor in the crystal coordinate that satisfies the requirements of symmetric transformation is given by

$$\overleftrightarrow{\sigma} = \begin{bmatrix} \sigma_\perp & 0 & 0 \\ 0 & \sigma_\perp & 0 \\ 0 & 0 & \sigma_\parallel \end{bmatrix} \quad (2)$$

In the laboratory frame (*x-y-z*), the crystal orientation is defined by the azimuthal angle $\varphi$ and the polar angle $\theta$. The conductivity tensor in the *x-y-z* coordinate is given by

$$\overleftrightarrow{\sigma}'_{xyz}(\theta,\varphi)$$

$$= \begin{bmatrix} \cos\varphi & -\sin\varphi & 0 \\ \sin\varphi & \cos\varphi & 0 \\ 0 & 0 & 1 \end{bmatrix} \begin{bmatrix} \cos\theta & 0 & -\sin\theta \\ 0 & 1 & 0 \\ \sin\theta & 0 & \cos\theta \end{bmatrix} \begin{bmatrix} \sigma_\perp & 0 & 0 \\ 0 & \sigma_\perp & 0 \\ 0 & 0 & \sigma_\parallel \end{bmatrix} \begin{bmatrix} \cos\theta & 0 & \sin\theta \\ 0 & 1 & 0 \\ -\sin\theta & 0 & \cos\theta \end{bmatrix} \begin{bmatrix} \cos\varphi & \sin\varphi & 0 \\ -\sin\varphi & \cos\varphi & 0 \\ 0 & 0 & 1 \end{bmatrix}$$

$$= \begin{bmatrix} (\sigma_\perp \cos^2\theta + \sigma_\parallel \sin^2\theta)\cos^2\varphi + \sigma_\perp \sin^2\varphi & (\sigma_\perp \cos^2\theta + \sigma_\parallel \sin^2\theta - \sigma_\perp)\sin\varphi\cos\varphi & (\sigma_\perp - \sigma_\parallel)\cos\theta\sin\theta\cos\varphi \\ (\sigma_\perp \cos^2\theta + \sigma_\parallel \sin^2\theta - \sigma_\perp)\cos\varphi\sin\varphi & (\sigma_\perp \cos^2\theta + \sigma_\parallel \sin^2\theta)\sin^2\varphi + \sigma_\perp \cos^2\varphi & (\sigma_\perp - \sigma_\parallel)\cos\theta\sin\theta\sin\varphi \\ (\sigma_\perp - \sigma_\parallel)\cos\theta\sin\theta\cos\varphi & (\sigma_\perp - \sigma_\parallel)\cos\theta\sin\theta\sin\varphi & \sigma_\perp \sin^2\theta + \sigma_\parallel \cos^2\theta \end{bmatrix}.$$

(3)

As a result, when charge currents are injected along the -*z* direction (electron currents $j_z^e$ along *z* in Fig. 1a), transverse currents along the *x*-and *y*-directions can be induced, with the conversion efficiency characterized by $\beta_0 = \frac{(\sigma_\perp - \sigma_\parallel)\cos\theta\sin\theta}{\sigma_\perp \sin^2\theta + \sigma_\parallel \cos^2\theta + (\sigma_\perp - \sigma_\parallel)\cos\theta\sin\theta}$. When $(\sigma_\perp - \sigma_\parallel)\cos\theta\sin\theta \ll \sigma_\perp \sin^2\theta + \sigma_\parallel \cos^2\theta$, we obtain $\beta_0 \approx \frac{(\sigma_\perp - \sigma_\parallel)\cos\theta\sin\theta}{\sigma_\perp \sin^2\theta + \sigma_\parallel \cos^2\theta}$.

## 3. Model for thickness dependence of terahertz amplitude.

To model the terahertz emission amplitude of the NM/EAC bilayer, we make use of Eq. (1). The impedance of the bilayer is given by

$$Z(\omega) = \frac{Z_0}{n_{air}(\omega) + n_{TiO_2}(\omega) + Z_0 \cdot \int_0^d dz \sigma(z,\omega)}, \quad (4)$$

where $\omega$ is the terahertz frequency, $Z_0 = 377\ \Omega$ is the vacuum impedance, $d = d_{NM} + d_{EAC}$ is the total thickness of the heterostructure, $\sigma(z,\omega)$ is the space-dependent sample conductivity, and $n_{air}$ and $n_{TiO_2}$ are the refractive indices at the terahertz frequency of air and TiO$_2$, respectively. As noted in Ref. 2, after excitation by the pump pulse, the terahertz signal is only contributed by the hot electrons injected from the NM layer that fall within the electron diffusion length of the EAC layer, $\lambda_{EAC}$. Similarly, only excited electrons within the diffusion length of the NM layer, $\lambda_{NM}$, are able to propagate through the interface without scattering. As a result, the terahertz radiation is caused by the charge currents localized near the interface of the heterostructure.

Following the above assumptions and the formalism in Ref. 50, we obtain the spatial distribution of the ballistic charge current density inside the EAC layer by

$$j_z(z) = j(d_{NM}) \frac{\sinh\left(\frac{z - d_{EAC}}{\lambda_{EAC}}\right)}{\sinh\left(\frac{d_{EAC}}{\lambda_{EAC}}\right)}, \quad (5)$$

and the injected charge current density $j(d_{NM})$ is proportional to the $z$-integration of the photoexcited hot electron density over the NM layer by considering $\lambda_{NM}$:

$$j(d_{NM}) \propto \frac{F_{abs}}{d_{NM} \hbar \omega_0} \tanh\left(\frac{d_{NM}}{2\lambda_{NM}}\right). \quad (6)$$

By inserting Eq. (4-6) into Eq. (1), we obtain

$$E_x(\omega) \propto \frac{F_{abs}}{d_{NM} \hbar \omega_0} \cdot \frac{Z_0}{n_{air}(\omega) + n_{TiO_2}(\omega) + Z_0 \cdot (\sigma_{EAC} d_{EAC} + \sigma_{NM} d_{NM})} \cdot \tanh\left(\frac{d_{NM}}{2\lambda_{NM}}\right) \cdot \tanh\left(\frac{d_{EAC}}{2\lambda_{EAC}}\right), \quad (7)$$

with the two diffusion lengths ($\lambda_{NM}$ and $\lambda_{EAC}$) and a global amplitude being the only free parameters. Here, $F_{abs}$ considers only the absorbed fluence by the NM layer, which is also thickness dependent by considering the reflection and absorption loss on the EAC layer and the NM layer:

$$F_{abs} = F \cdot (1 - R_1) \cdot (1 - R_2) \cdot e^{-\alpha_{NM} d_{NM}} \cdot \left(1 - e^{-\alpha_{EAC} d_{EAC}}\right), \tag{8}$$

where $R_1$ and $R_2$ are the reflectivity of the EAC-air interface and the EAC-NM interface, respectively, $\alpha_{NM}$ and $\alpha_{EAC}$ are the optical absorption coefficients of the NM and the EAC materials, respectively. Most of the optical and electrical parameters in the model can be determined by the literature values (see SM Section S3) or by experimental measurements, leaving the hot-electron velocity relaxation lengths[2] ($\lambda_{NM}$ and $\lambda_{EAC}$) and a global amplitude as the only free parameters. The fitting to the experimental results in Fig. 4b and c yields $\lambda_{NM} \approx 1$ nm for Pt and $\lambda_{EAC} \approx 3.2$ nm for RuO$_2$. The former is consistent with previous work[2].

**Acknowledgement**

This work was accomplished at Fudan University. L. Z. and Z. T. acknowledge the support from the National Key Research and Development Program of China (Grant No. 2022YFA1404700). C. T. and Z. T. also acknowledge the support from the National Key Research and Development Program of China (Grant No. 2021YFA1400200). L. Z., C. T., Y. W. and Z. T. acknowledge the support from the National Natural Science Foundation of China (Grant No. 12221004). Y. W. and Z. T. acknowledge the support from the Shanghai Municipal Science and Technology Basic Research Project (Grant No. 22JC1400200). Y. W. acknowledges the support from the National Key Research Program of China (Grant No. 2022YFA1403300), the National Natural Science Foundation of China (Grants No. 11974079, No. 12274083) and the Shanghai Municipal Science and Technology Major Project (Grant No. 2019SHZDZX01). L. Z. acknowledges the support from Science Foundation of Shanghai (Grant No. 20JC1414601). Z. Y. acknowledges the support from National Natural Science Foundation of China (No. 12174028). Z. T. acknowledges the support from the National Natural Science Foundation of China (No. 12274091).

**Table 1.** Longitudinal ($\sigma_\parallel$) and transverse ($\sigma_\perp$) conductivities, crystal polar angles $\theta$ and the coefficients of electrical anisotropy $\beta_0$ of RuO$_2$(101) and IrO$_2$(101).

|  | $\sigma_\parallel$ ($\times 10^5 \Omega^{-1} m^{-1}$) | $\sigma_\perp$ ($\times 10^5 \Omega^{-1} m^{-1}$) | $\theta$ | $\beta_0$ |
|---|---|---|---|---|
| RuO$_2$(101) | 8.13 | 8.63 | 34.7 | 0.028 |
| IrO$_2$(101) | 3.18 | 4.22 | 35.0 | 0.139 |

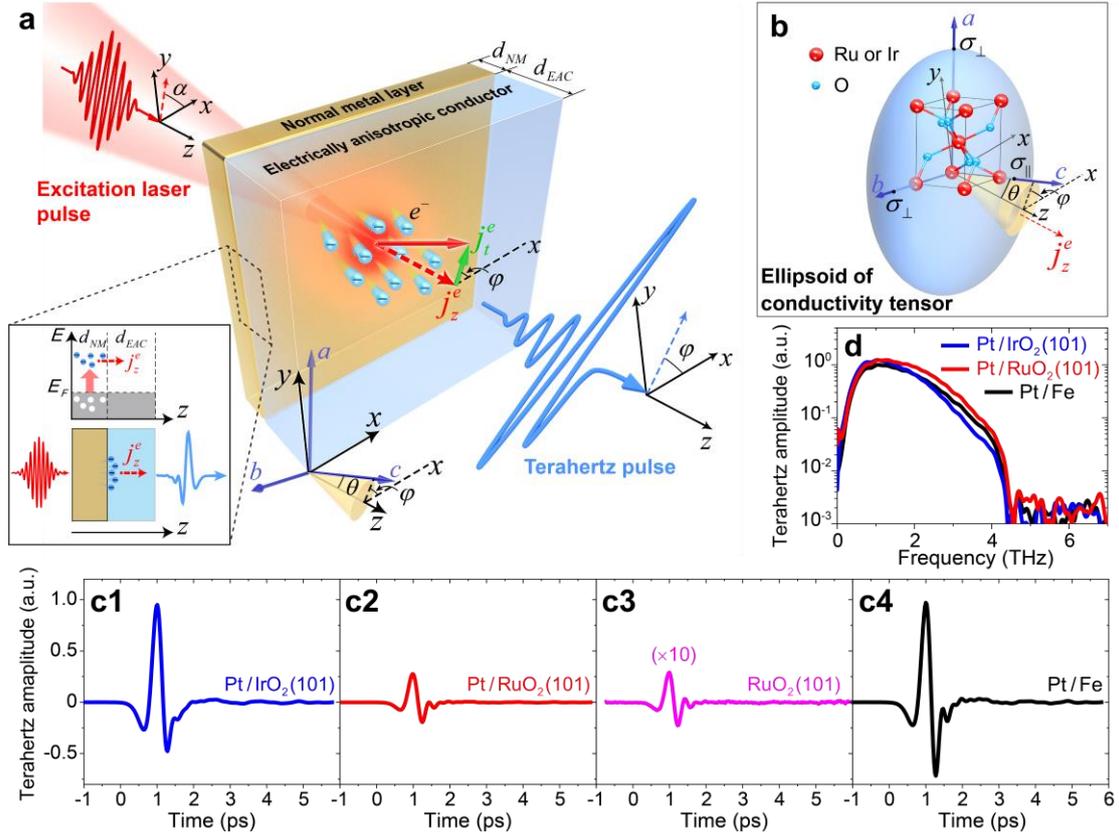

**Figure 1. Experimental setup and terahertz signals. a.** Schematic of the experimental setup. The *x-y-z* coordinate is adapted to the laboratory frame and the lattice coordinates are labeled as *a-b-c*. The crystal azimuthal angle $\varphi$, polar angle $\theta$ and the laser-polarization angle $\alpha$ are defined. Femtosecond-laser induced electrons are injected from the NM layer, resulting in transient electron currents $j_z^e$ along *z*. In the experiment, the excitation pulse is primarily incident from the EAC layer side. However, for ease of illustration, it is depicted as incident from the NM layer side. Owing to the electrical anisotropy in the EAC layer, transverse electron currents ($j_t^e$) are generated flowing at an angle of $\varphi$ relative to the *x*-axis. **Inset:** Illustration of super-diffusive electron currents $j_z^e$ induced by laser-pulse excitation. **b.** Schematic of the ellipsoid of conductivity tensor of $RuO_2$ and $IrO_2$ in the laboratory frame, displaying anisotropy in electrical conductivities ($\sigma_\parallel < \sigma_\perp$). The schematic uses the same coordinate and angle definitions as described in **a**. **c.** Terahertz waveforms generated from $Pt/IrO_2(101)$ (**c1**), $Pt/RuO_2(101)$ (**c2**), $RuO_2(101)$ (**c3**), and $Pt/Fe$ (**c4**) devices. The signal from the $RuO_2(101)$ thin film is scaled 10 times for comparison. **d.** Terahertz spectra obtained via fast Fourier transform (FFT) of the waveforms in **c**.

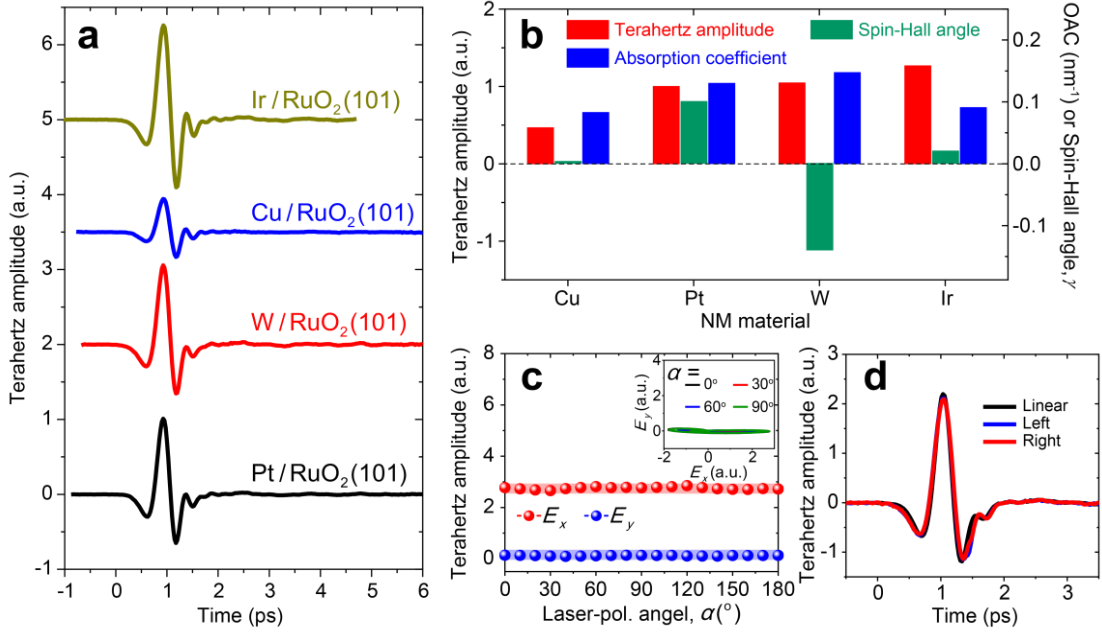

**Figure 2. Effect of NM materials and laser polarization states. a.** Terahertz waveforms generated by NM/RuO$_2$(101) devices with NM materials of Ir, Cu, W and Pt. The thickness of the NM layer is 2 nm. **b.** Terahertz signal amplitude as a function of the NM materials used for the NM/RuO$_2$(101) devices (red bars). For comparison, optical absorption coefficients (OAC) at the laser wavelength of 1.03 μm (blue bars) and spin-Hall angles γ (green bars) of the respective NM materials are also shown. **c.** Terahertz signal amplitudes of the $E_x$ and $E_y$ components from the Pt/RuO$_2$(101) device as a function of the polarization angle α of the linearly polarized excitation laser. **Inset:** $E_x$-$E_y$ projection of the terahertz waves for different values of α. **d.** Terahertz waveforms from the Pt/RuO$_2$(101) device excited by excitation pulses with linear, right- and left-circular polarizations.

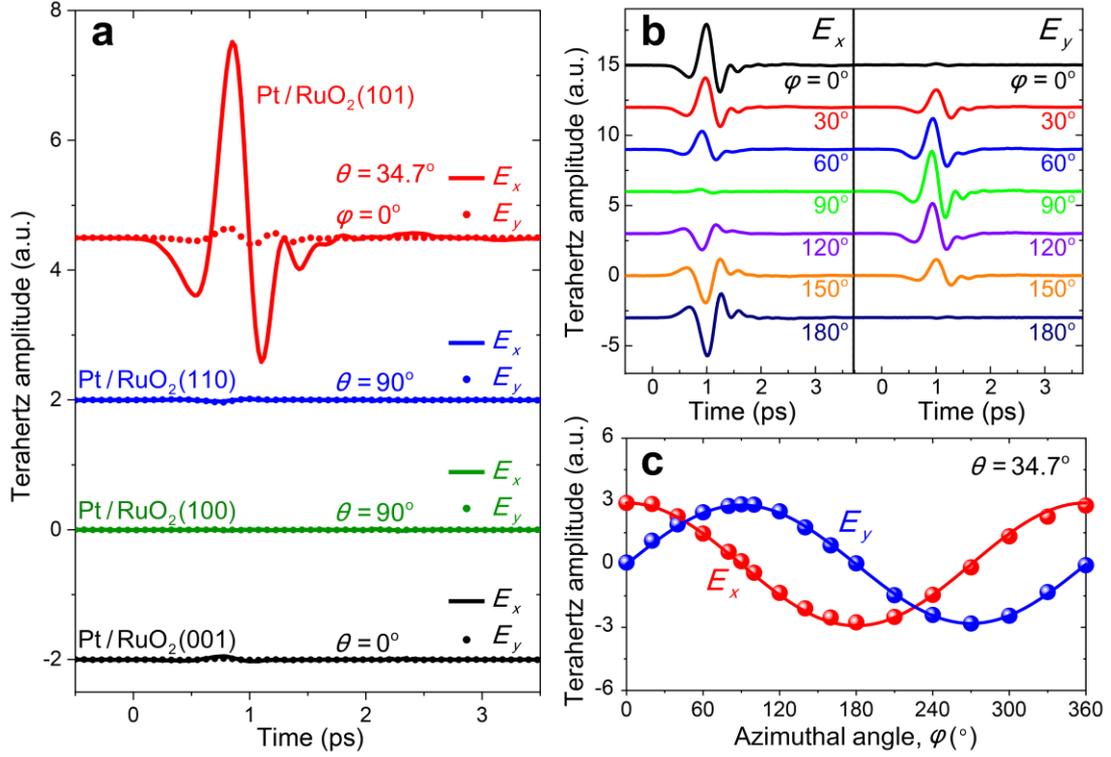

**Figure 3. Effect of crystal orientations. a.** $E_x$ and $E_y$ components of terahertz waveforms generated by NM/RuO$_2$ devices with different crystal orientations with different polar angles $\theta$. **b.** $E_x$ and $E_y$ components of terahertz waveforms generated by Pt/RuO$_2$(101) at different azimuthal angle $\varphi$ while keeping $\theta$ fixed at 34.7°. **c.** Terahertz signal amplitude of $E_x$ and $E_y$ components from the Pt/RuO$_2$(101) device at different azimuthal angles $\varphi$. The solid lines represent the sine and cosine fitting to the results.

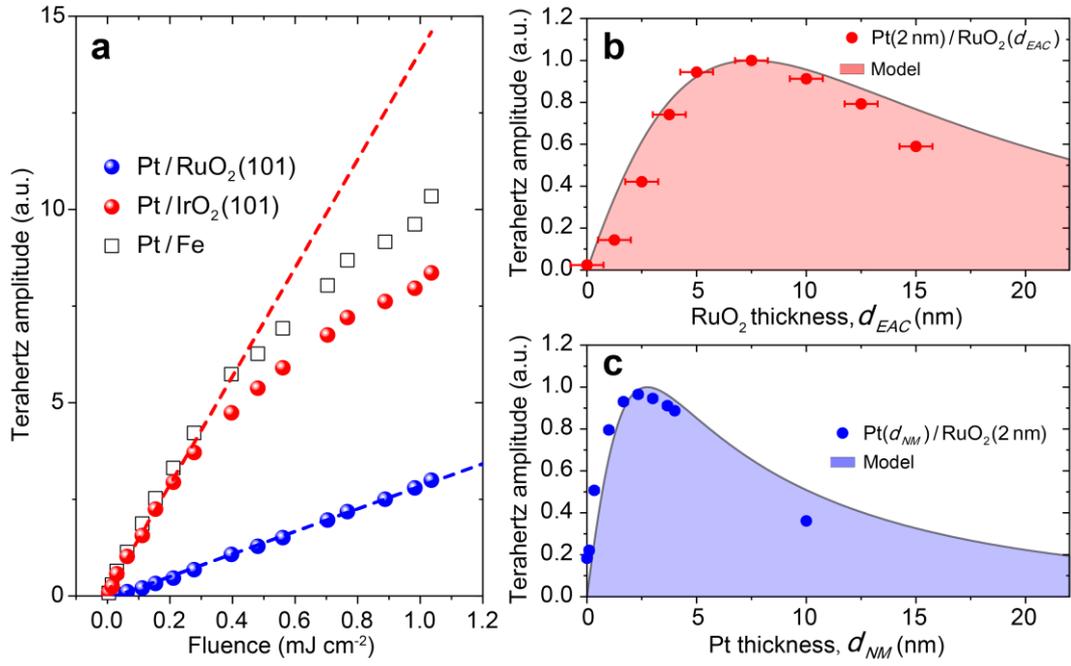

**Figure 4. Optimizing the conversion efficiency. a.** Terahertz signal amplitude as a function of incident laser fluence from the Pt/RuO$_2$(101), Pt/IrO$_2$(101) and Pt/Fe samples. The red and blue dashed lines represent the linear fits to the low-fluence experimental results of Pt/IrO$_2$(101) and Pt/RuO$_2$(101), respectively. The slope of the red-dashed line is approximately 4.8 times of that of the blue-dashed line. **b.** Terahertz signal amplitude as a function of thickness of the RuO$_2$ layer ($d_{EAC}$) of the Pt/RuO$_2$(101) device. **c.** Terahertz signal amplitude as a function of thickness of the Pt layer ($d_{NM}$) of the Pt/RuO$_2$(101) device. The solid lines in **b** and **c** represent a global fit using the thickness-dependent model (see Methods).